\pdfoutput = 1

\documentclass[12pt,a4paper]{article}

\usepackage{cite}
\usepackage{placeins}

\usepackage[utf8]{inputenc}

\usepackage{amsmath}
\usepackage[dvipsnames]{xcolor}
\usepackage{amssymb}
\usepackage{graphicx}
\usepackage[colorlinks=true, allcolors=Blue]{hyperref}
\usepackage{caption}
\usepackage{subcaption}
\usepackage{braket}
\usepackage{slashed}
\usepackage{float}
\usepackage{multirow}
\usepackage{cite}

\mathcode`l="8000
\begingroup
\makeatletter
\lccode`\~=`\l
\DeclareMathSymbol{\lsb@l}{\mathalpha}{letters}{`l}
\lowercase{\gdef~{\ifnum\the\mathgroup=\m@ne \ell \else \lsb@l \fi}}%
\endgroup
\usepackage{xspace}

\newcommand{\Bbar}{\kern 0.18em\overline{\kern -0.18em B}{}\xspace}

\newcommand{\Kbar}{\kern 0.18em\overline{\kern -0.18em K}{}\xspace}

\begin{document}
\begin{titlepage}

\vspace*{-0.7truecm}
\begin{flushright}
Nikhef-2024-003
\end{flushright}

\vspace{1.6truecm}

\begin{center}
\boldmath
{\Large{\bf CP Violation in $B$ Decays:  Recent Developments \\

\vspace*{0.3truecm}

and Future Perspectives}}
\unboldmath
\end{center}

\vspace{0.8truecm}

\begin{center}
{\bf Robert Fleischer\,${}^{a,b}$}

\vspace{0.5truecm}

${}^a${\sl Nikhef, Science Park 105, NL-1098 XG Amsterdam, Netherlands}

${}^b${\sl  Department of Physics and Astronomy, Vrije Universiteit Amsterdam,\\
NL-1081 HV Amsterdam, Netherlands}

\end{center}

\vspace*{1.7cm}

\begin{abstract}
\noindent
CP violation in B decays provides a powerful tool to probe physics from beyond the Standard Model. A theoretical overview of recent developments of benchmark channels is given, ranging from non-leptonic to rare leptonic and semileptonic modes, opening up exciting perspectives for the 
future high-precision era of flavour physics and the pursuit of New Physics.
\end{abstract}

\vspace*{2.1truecm}

\noindent
{\it Invited contribution to EPJ ST Special Issue: b-quark physics as a precision laboratory: A collection of articles on the present status and prospects for the future, Guest Editors: Rusa Mandal, B. Ananthanarayan, and Daniel Wyler}

\vfill

\noindent
January 2024

\end{titlepage}

\newpage

\thispagestyle{empty}
\vbox{}
\newpage

\setcounter{page}{1}

\section{Introduction}\label{sec:intro}
CP violation offers powerful probes for testing the Standard Model (SM) and searching for footprints of New Physics (NP). In this endeavour, decays of $B$ mesons play an outstanding role. In weak $B$ decays, CP-violating asymmetries may originate through subtle interference effects. Since we have to deal with hadronic bound states in these phenomena, strong interactions play a key role and may also provide an important ingredient to CP asymmetries through CP-conserving strong phases. However, these effects represent also a major challenge for theoretical analyses due to long-distance effects which cannot be analysed in perturbation theory. Fortunately, the $B$-meson system provides a variety of decays allowing strategies to deal with these challenges: we may either have decays, where hadronic matrix elements cancel in CP-violating asymmetries or information from further decay channels can be utilised to determine and constrain strong interaction effects through experimental data. The key players in this field are non-leptonic decays. However, also rare decays into leptonic or semileptonic final states offer exciting probes for physics beyond the SM through CP-violating phenomena.

The quark flavour sector of the SM is encoded in the Cabibbo--Kobayashi--Maskawa (CKM) matrix \cite{cab,KM}. This quark-mixing matrix gives rise to a plethora of flavour-physics phenomena. The corresponding ``quark-flavour code" is governing weak decays of $K$, $D$ and $B$ mesons. CP violation can be 
accommodated through a complex phase in the CKM matrix. Despite the tremendous success of the SM, we have indications that this framework cannot be complete. Key examples are dark matter and the non-vanishing neutrino masses. Particularly intriguing is the baryon asymmetry of the Universe, which suggests that the CP violation present in the SM is too small by many orders of magnitude for generating this asymmetry. Consequently, we are missing sources of CP violation.
When going beyond the SM, typically new sources for flavour and CP violation arise. It is also important to note that we still do not understand the origin of the structure of the SM and the patterns of masses and flavour-mixing parameters with their intriguing hierarchies at a more fundamental level. 

In order to search for new particles, the obvious approach is to try to produce them in particle collisions and detect their decay products in dedicated detectors. Here the limiting key factor is the centre-of-mass energy of the collider. Data taken 
through proton--proton collisions at the Large Hadron Collider (LHC) have led to the discovery of the Higgs boson by the ATLAS and CMS collaborations in 2012 \cite{ATLAS-Higgs,CMS-Higgs}. These experiments are further exploring the features and decay properties of this fascinating particle, which looks so far like the scalar boson arising in the minimal Higgs mechanism of the Standard Model to break the electroweak symmetry. However, no other new particles were seen by these experiments so far.

Instead of searching for new particles at the high-energy frontier, we may utilise possible imprints of new particles arising from quantum fluctuations in rare processes. Here the key point is that new interactions and the associated new particles may manifest themselves through contributions to decay observables, thereby resulting in discrepancies between the corresponding SM predictions and measurements. In these explorations, precision -- both theoretical and experimental -- is the key limitation. New physics particles much heavier than those that could be produced at colliders could be revealed in such studies. In the future era of particle physics, this high-precision frontier will plays a key role. Decays of $B$ mesons and CP-violating phenomena are particularly promising in this respect. Interestingly, the current data show puzzling patterns for various $B$ decays although the situation is unfortunately not (yet) conclusive. 

How can we summarise this situation? We may conclude that NP entering a Lagrangian
\begin{equation}
{\cal L}={\cal L}_{\rm SM} + {\cal L}_{\rm NP}(m_{\rm NP}, g_{\rm NP}, \varphi_{\rm NP},  ...)
\end{equation}
through $ {\cal L}_{\rm NP}$ with parameters such as masses $m_{\rm NP}$, couplings $g_{\rm NP}$ and CP-violating phases $\varphi_{\rm NP}$
is characterised by a large NP scale $\Lambda_{\rm NP}$ that is far beyond the TeV regime probed by the LHC, 
which would be challenging for direct searches at  ATLAS and CMS, or/and symmetries prevent large NP effects in flavour-changing neutral currents and the flavour sector. Fortunately, we are facing exciting future prospects, in particular due to the LHC upgrade(s) and the data taking at the Belle II experiment. 

A powerful theoretical framework to deal with NP effects entering far beyond the electroweak scale is given by ``Effective Field Theories". Here 
 the heavy degrees of freedom, i.e.\ the NP particles as well as the top quark and the $Z$ and $W$ bosons of the SM, are ``integrated out" from appearing explicitly in analyses of low-energy phenomena, such as weak decays of $B$ mesons. Their effects are described by short-distance functions, which may also  have complex phases arising from new sources of CP violation. In recent decades, perturbative QCD corrections were calculated and renormalisation group techniques applied for the summation of large logarithms. Such analyses have been performed for the SM, but also a wide spectrum of specific NP scenarios. In order to perform model-independent analyses of NP effects, Standard Model Effective Theories (SMEFT) offer an interesting tool.

Let us have a closer look at this formalism for $B$-meson decays. We may calculate the  $\bar{B}\to \bar{f}$ transition amplitude as the 
matrix element of the corresponding low-energy effective Hamiltonian, resulting in the following general structure within the SM \cite{BBL}:
\begin{equation}
\langle \bar{f}|{\cal H}_{\mbox{{\scriptsize eff}}}|\bar{B}\rangle=
\frac{G_{\rm F}}{\sqrt{2}}\sum_{j}\lambda_{\rm CKM}^{j}\sum_{k}
C_{k}(\mu) \langle \bar{f}|Q^{j}_{k}(\mu)|\bar{B}\rangle.
\end{equation}
Here $G_{\rm F}$ is the Fermi constant, the $\lambda_{\rm CKM}^j$ denote combinations of CKM matrix elements relevant for the considered decay class, 
$\mu$ is a renormalisation scale and the $Q^{j}_{k}$ are four-fermion operators with their short-distance Wilson coefficients $C_{k}(\mu)$. Such 
Hamiltonians characterise different quark-flavour processes, and specific decays are described through the corresponding hadronic matrix 
elements. Whereas the Wilson coefficients can be calculated in perturbation theory, including QCD corrections leading to the $\mu$ dependences, 
the non-perturbative hadronic matrix elements are usually associated with hadronic uncertainties. The $\mu$ dependence of the short-distance coefficients
is cancelled by that of the hadronic matrix elements, thereby resulting in $\mu$-independent transition amplitudes. This feature gives rise to renormalisation group equations, allowing the summation of large logarithms.

When going beyond the SM, the Wilson coefficients of SM operators may get new contributions. Moreover, new operators -- absent in the SM -- may arise that could be associated with new sources of CP violation, which are encoded as complex phases of their coefficients. If we consider a specific NP model, the corresponding low-energy effective Hamiltonian could be calculated and the short-distance functions be expressed in terms of the parameters of the model. Conversely, we may also just add all operators which could contribute to a given decay. The latter approach is particularly interesting as it allows model-independent analyses of NP effects. Using experimental data, the short-distance coefficients can be constrained or even determined. In order to narrow down the underlying NP, correlations between observables play an essential role. Should it be possible to eventually establish NP effects, the next goal would be to go beyond the EFT description and to construct a ``New Standard Model". 

Non-leptonic $B$ decays with only hadrons in the final states are the most challenging processes with respect to the impact of strong interactions. 
These decays are key probes for the exploration of CP violation. 
The reason is that non-vanishing CP-violating rate asymmetries require interference effects which may arise in such channels in various ways. 
Theoretical predictions are affected by hadronic matrix elements of four-quark operators. Fortunately, it is possible to circumvent the calculation of these non-perturbative quantities in studies of CP violation: 
\begin{itemize}
\item Amplitude relations can be utilised to eliminate hadronic matrix elements and/or to determine them through experimental data. Nature provides exact relations, involving pure ``tree" decays of the kind $B\to D K$, as well as approximate relations following from the flavour symmetries of strong interactions, i.e.\ $SU(2)$ isospin or flavour $SU(3)$, involving channels of the kind $B\to\pi\pi$, $B\to\pi K$ and $B_{(s)}\to KK$.
\item In decays of neutral mesons, interference effects may arise from $B^0_q$--$\bar{B}^0_q$ ($q=d,s$) mixing should both the $B^0_q$ and 
its anti-particle 
decay into the same final state. If one CKM amplitude dominates the decay, the corresponding hadronic matrix elements cancel in such ``mixing-induced" CP asymmetries, while ``direct" CP violation -- arising directly at the decay amplitude level through interference between different decay contributions -- would vanish.
\end{itemize}
Measurements of CP violation in non-leptonic $B$ decays allow determinations of the angles of the Unitarity Triangle (UT) of the CKM matrix, thereby playing a key role in the testing of the SM. For detailed UT analyses, the reader is referred to Refs.~\cite{CKMfitter,UTfit}. In the presence of CP-violating NP contributions, discrepancies should emerge between constraints from various processes. 

In our quest for physics beyond the SM, we are moving towards new frontiers. For resolving potentially small NP effects, it is crucial to have a critical 
look at theoretical SM analyses and their approximations, where strong interactions lead to hadronic uncertainties. The goal is to match 
the experimental and theoretical precisions, pushing both to the same level. Let us in the following section have a first closer look at this 
challenge for benchmark channels of CP violation.

\boldmath
\section{The $B^0_d\to J/\psi K_{\rm S}$ and $B^0_s\to J/\psi \phi$ Decays}\label{sec:BJpsi}
\unboldmath
The $B^0_d\to J/\psi K_{\rm S}$ channel \cite{CaSa1,CaSa2,BiSa} and its counterpart $B^0_s\to J/\psi \phi$ \cite{DDF,DFN}, where the down spectator quark is replaced by a strange quark, belong to the most prominent channels for exploring CP violation. The $B^0_d\to J/\psi K_{\rm S}$ decay 
has led to the observation of CP violation in the $B$ system 
by the BaBar \cite{BaBar-CP-obs} and Belle \cite{Belle-CP-obs} collaborations in 2001. In the SM, the $B_d^0\to J/\psi\, K_{\rm S}$ decay amplitude can be written as follows \cite{RF-99}:
\begin{equation}\label{ABdpsiKS}
A(B_d^0\to J/\psi\, K_{\rm S})=\left(1-\lambda^2/2\right){\cal A'}
\left[1+ \epsilon a'e^{i\theta'}e^{i\gamma}\right].
\end{equation}
Here the hadronic parameters ${\cal A'}$ and $a'e^{i\theta'}$ describe colour-suppressed tree and penguin topologies, respectively, $\gamma$ is the usual angle of the UT, $\lambda\equiv |V_{us}|\approx 0.22$, and 
\begin{equation}\label{eps-def}
\epsilon\equiv\frac{\lambda^2}{1-\lambda^2}\approx0.05
\end{equation} 
is a doubly Cabibbo-suppressed CKM parameter. 

Concerning the CP-violating asymmetries, let us consider a neutral $B_q$ meson ($q=d,s$) decaying into a final state with CP 
eigenvalue $\eta_f$. Due to the
$B^0_q$--$\bar B^0_q$ oscillations and the associated time evolution of initially present $\bar B_q^0$ or $B_q^0$ states, we obtain the 
following time-dependent CP asymmetry:
\begin{eqnarray}
    a_{\text{CP}}(B_q(t)\to f)  & \equiv &
    \frac{|A(B_q^0(t)\to f)|^2-|A(\bar B_q^0(t)\to f)|^2}{|A(B_q^0(t)\to f)|^2+|A(\bar B_q^0(t)\to f)|^2} \nonumber \\
    & = & \frac{\mathcal{A}_{\text{CP}}^{\text{dir}}(B_q\to f)\cos(\Delta m_qt)+\mathcal{A}_{\text{CP}}^{\text{mix}}(B_q\to f)\sin(\Delta m_qt)}
    {\cosh(\Delta\Gamma_qt/2)+\mathcal{A}_{\Delta\Gamma}(B_q\to f)\sinh(\Delta\Gamma_qt/2)},\label{CP-asym-TD}
    \end{eqnarray} 
where $\Delta m_q\equiv m^{(q)}_{\text{H}}-m^{(q)}_{\text{L}}$ and $\Delta\Gamma_q\equiv \Gamma_{\text{L}}^{(q)}-\Gamma_{\text{H}}^{(q)}$ are the mass and decay width difference between the heavy and light eigenstates of the $B_q$-meson system, respectively. While the decay width difference is negligibly small in the $B_d$ system, it is sizeable for $B_s$ mesons and characterised by the parameter 
$y_s\equiv \Delta\Gamma_s/(2\,\Gamma_s)=0.062 \pm 0.004$. The direct CP asymmetry 
$\mathcal{A}_{\text{CP}}^{\text{dir}}$ originates from interference between different amplitudes with both non-trivial CP-violating phases (CKM elements) and non-trivial CP-conserving phases (strong rescattering effects). On the other hand, the mixing-induced CP asymmetry $\mathcal{A}_{\text{CP}}^{\text{mix}}$ is generated through interference between the $\bar B_q^0 \to f$ and $B_q^0\to f$ decay amplitudes induced by the neutral $B_q$-meson 
oscillations. It should be noted that the observables satisfy the following general relation:
\begin{equation}\label{cons-rel}
\mathcal{A}_{\text{CP}}^{\text{dir}}(B_q\to f)^2 + \mathcal{A}_{\text{CP}}^{\text{mix}}(B_q\to f)^2 +\mathcal{A}_{\Delta\Gamma}(B_q\to f)^2 =1\,.
\end{equation}

In the SM, $B^0_q$--$\bar{B}^0_q$ mixing arises from box topologies. This phenomenon is very sensitive to possible NP contributions which may either enter the loop processes or at the tree level, as, for instance, in models with extra $Z'$ bosons. In general, such effects involve also new sources for CP violation. The CP-violating phases associated with $B^0_q$--$\bar{B}^0_q$ mixing can be written as follows:
\begin{equation}\label{phi-q}
\phi_d=2\beta+\phi_d^{\rm NP}, \quad \phi_s=-2\lambda^2\eta +\phi_s^{\rm NP},
\end{equation}
where $\beta$ is the usual angle of the UT, $\lambda\equiv |V_{us}|\approx 0.22$ and $\eta$ are CKM parameters, and the 
$\phi_q^{\rm NP}$ denote possible CP-violating NP phases. In order to quantify NP effects, the following model-independent parametrisation can be introduced \cite{BF-NP,Bq-NP-analysis}:
\begin{equation}\label{NP-Bq-par}
    \Delta m_{q}  = \Delta m_{q}^{\text{SM}} \left(1 + \kappa_{q} e^{i\sigma_{q}}\right), \quad
    \phi_{q}^{\text{NP}}  =  \arg\left(1 + \kappa_{q} e^{i\sigma_{q}}\right), 
\end{equation}
where $\kappa_q$ measures the size of the NP effects with respect to the SM and  $\sigma_q$ describes a CP-violating phase which 
is induced by new sources of CP violation. 

The final state of the $B^0_d\to J/\psi K_{\rm S}$ decay is a CP eigenstate with eigenvalue $\eta_{J/\psi K_{\rm S}}=-1$. Here it is assumed that the $K_{S}$ is a CP-even eigenstate, neglecting tiny CP violation at the $10^{-3}$ level of the neutral kaon system (for a recent review of CP violation in the kaon system, see Ref.~\cite{AJB-kaon}). The direct and mixing-induced CP asymmetries of the $B^0_d\to J/\psi K_{\rm S}$  channel satisfy the following relation:
\begin{equation}\label{CP-rel-1}
\frac{\eta_{J/\psi K_{\rm S}}\mathcal{A}_{\text{CP}}^{\text{mix}}(B_d\to J/\psi K_{\rm S})}{\sqrt{1-\mathcal{A}_{\text{CP}}^{\text{dir}}(B_d\to J/\psi K_{\rm S})^2}}
=\sin(\phi_d+\Delta\phi_d),
\end{equation}
where $\Delta\phi_d$ is a hadronic phase shift \cite{FFJM}:
\begin{equation}
\sin\Delta\phi_d \propto 2 \epsilon a'\cos\theta' \sin\gamma+\epsilon^2a'^2, \quad
\cos\Delta\phi_d \propto 1+ 2 \epsilon a'\cos\theta' \cos\gamma+\epsilon^2a'^2
\cos2\gamma.
\end{equation}
Neglecting the doubly Cabibbo-suppressed $\epsilon a'$ penguin parameters, we obtain the well-known result 
$\eta_{J/\psi K_{\rm S}}\mathcal{A}_{\text{CP}}^{\text{mix}}(B_d\to J/\psi K_{\rm S})=\sin\phi_d$, 
which is usually applied to determine the UT angle $\beta$ from the measured CP-violating asymmetries.

The decay $B^0_s\to J/\psi\phi$ is the $B^0_s$ counterpart of $B_d^0\to J/\psi\, K_{\rm S}$ and 
arises from the same quark-level processes. However, 
the final state with two vector mesons is a mixture of CP-odd and CP-even eigenstates $f=0,\parallel$ and $\perp$, respectively.
In order to disentangle them, an angular analysis of the $J/\psi \to\mu^+\mu^-$, $\phi \to\ K^+K^-$ decay products 
has to be performed in the time-dependent decay rate analysis \cite{DDF,DFN,dBF}. In analogy to $B_d^0\to J/\psi\, K_{\rm S}$, doubly Cabibbo-suppressed penguins effects lead to an effective CP-violating mixing phase entering the CP-violating 
observables \cite{FFM}:
\begin{equation}
\phi_{s,(\psi \phi)_f}^{\rm eff} = \phi_s+\Delta\phi_s^{(\psi \phi)_f} \equiv  \phi_s+\Delta\phi_s^{f}.
\end{equation}
For a small mixing phase $\phi_s$ in the few degree regime, as follows from the picture of the experimental data, even a 
hadronic phase shift $\Delta\phi_s^{f}$ at the $1^\circ$ level would have a significant impact. 

In the future high-precision era of Belle II and the LHCb upgrade(s), the experimental precision requires the control of the penguin corrections to 
reveal possible CP-violating NP contributions to $B^0_q$--$\bar B^0_q$ mixing. The topic receives long-standing interest in the 
theory community (see, e.g., Refs.~\cite{RF-99,FFJM,FFM,CPS,GR-08,jung,FNW,dBF,BdBFM}). Unfortunately, we cannot reliably calculate the hadronic phase shifts $\Delta\phi_d$ and $\Delta \phi_s^f$ within QCD from first principles. However, we may 
use ``control channels" to constrain and determine these effects with the help of flavour symmetries and experimental data. Key channels 
are $B^0_s\to J/\psi K_{\rm S}$, $B^0_d\to J/\psi \pi^0$, and $B^0_s\to J/\psi \rho^0$, which have a different CKM amplitude structure with the feature that 
the penguin parameters are not doubly Cabibbo-suppressed by the parameter $\epsilon$. 

Let us have a closer look at the $B^0_s\to J/\psi K_{\rm S}$ decay. The corresponding transition amplitude takes the following form \cite{RF-99}:
\begin{equation}
A(B^0_s\to J/\psi K_{\rm S})\propto\left[1-a e^{i\theta}e^{i\gamma}\right],
\end{equation}
which should be compared with Eq.~(\ref{ABdpsiKS}). We observe that the penguin parameter $a e^{i\theta}e^{i\gamma}$ does indeed not enter with $\epsilon$. If the CP asymmetries of this channel are measured, we may determine $a$ and $\theta$, which can then be related to their 
$B^0_d\to J/\psi K_{\rm S}$ counterparts through the $U$-spin symmetry of strong interactions,
\begin{equation}
a e^{i\theta}=a' e^{i\theta'}, 
\end{equation}
thereby allowing us to include their effects in the determination of $\phi_d$ from Eq.~(\ref{CP-rel-1}).

\begin{figure}[t] 
   \centering
   \includegraphics[width=8.0truecm]{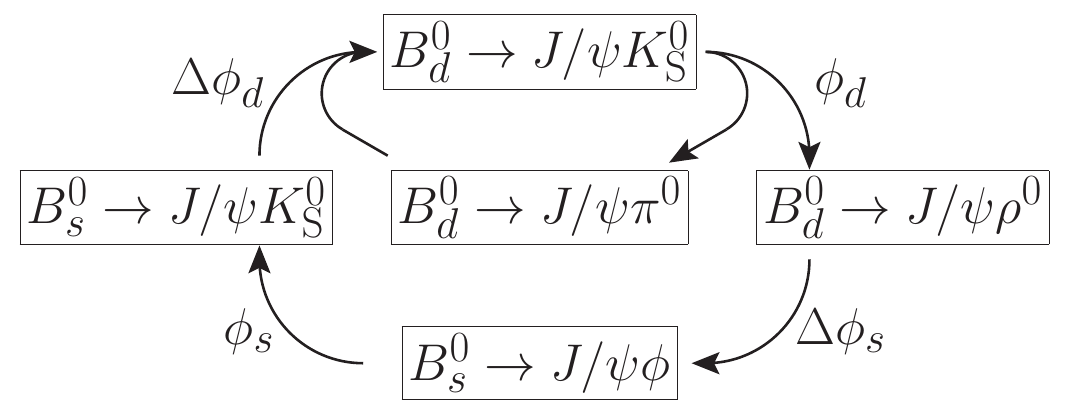} 
   \caption{Illustration of the interplay between the determination of the CP-violating phases $\phi_d$ and $\phi_s$ and their hadronic penguin phase
   shifts $\Delta\phi_d$ and $\Delta\phi_s$ through control channels (from Ref.~\cite{BdBFM}).}
   \label{fig:interplay}
\end{figure}

In Ref.~\cite{BdBFM}, a simultaneous strategy of various control channels was proposed and applied to the currently available data, utilising the $SU(3)$ flavour symmetry of strong interactions. The point is that there is a subtle interplay between the mixing phases and decays, as illustrated in Fig.~\ref{fig:interplay}. For a detailed discussion, also of the numerical analysis with correlation plots, the reader is referred to that paper and the more 
recent update given in Ref.~\cite{BdBFM-CKM}. Let us here just give the following main numerical results:
\begin{equation}\label{phi_d-res}
a=0.14^{+0.17}_{-0.11}, \quad \theta=\left(173^{+35}_{-45} \right)^\circ,  \quad
\phi_d=\left(44.4^{+1.6}_{-1.5} \right)^\circ,
\end{equation}
which should be compared with the measured value  $\phi_{d,J/\psi K^0}^{\rm eff}=\left(43.6\pm1.4\right)^\circ$, and
\begin{equation}\label{BpsiV}
a_V=0.044^{+0.0.085}_{-0.038}, \quad \theta_V=\left(306^{+48}_{-112} \right)^\circ,  \quad
\phi_s=-\left(4.2\pm1.4 \right)^\circ,
\end{equation}
which should be compared with $\phi_{s,J/\psi\phi}^{\rm eff}=-\left(4.1\pm1.3\right)^\circ$.

In the analysis of the 
$B_{(s)}\to J/\psi V$ modes, polarisation-dependent effects had to be ignored due to the current lack of data. In the future, it would be important to make polarisation-dependent measurements, which could then be implemented in the strategy to further refine the analysis. The results in Eq.\ (\ref{BpsiV})
correspond to the penguin shift $\Delta\phi_s= \left(0.14^{+0.54}_{-0.70}\right)^\circ$. Using Eq.~(\ref{phi-q}) with 
$\phi_s^{\rm SM}=-2\lambda^2\eta=-\left(2.01\pm0.12 \right)^\circ$ yields $\phi_s^{\rm NP}=-\left(2.2\pm1.4\right)^\circ$. The future scenarios studied in Ref.~\cite{BdBFM-CKM} show that CP-violating NP effects could be revealed in the high-precision era 
with $5\sigma$ significance. On the other hand, for the $B_d$-meson system, the SM prediction of $\phi_d$ is a limiting factor, thereby
making $\phi_d$ less favourable. 

A detailed analysis of possible NP contributions to $B^0_q$--$\bar{B}^0_q$ mixing, as described by Eq.\ (\ref{NP-Bq-par}), was performed in 
Ref.~\cite{Bq-NP-analysis}, utilising the results for the mixing phase extracted from the experimental data discussed above. For a selection of earlier analyses, see Refs.\ \cite{UT-NP,LN-NP-1,LN-NP-2,C-CKM-NP}. In order to constrain the NP parameters $\kappa_q$ and $\sigma_q$, the SM predictions for the $B_q$ mixing parameters play the central role. In order to calculate them, the apex of the UT has to constrained through the angle $\gamma$ and the CKM matrix elements $|V_{cb}|$ and $|V_{ub}|$. Unfortunately, concerning the determination of the latter parameters from semileptonic $B$ decays, we are facing long-standing discrepancies between extractions utilising inclusive and exclusive 
channels \cite{KV-Beauty-23}. 
Special care is needed in view of this unsatisfactory situation, which will hopefully be resolved in the future high-precision era in particular thanks to Belle II.
In Ref.~\cite{Bq-NP-analysis}, various determinations of the UT and constraints for $(\kappa_q,\sigma_q)$ were obtained, addressing these issues in detail.
In the case of the $B_d$-meson system, the SM reference value of $\phi_d$ turns out to be the major limiting factor, while the $B_s$ system is much more favourable in this respect. An interesting application of the NP parameters $(\kappa_q,\sigma_q)$ concerns the prediction of the branching ratio of the rare
decay $B^0_s\to \mu^+\mu^-$: As noted in Refs.\ \cite{AJB-03,Bo-Bu,Bu-Ve}, in the ratio with the mass difference $\Delta M_s$, the 
CKM matrix element $|V_{cb}|$ cancels. The information for the NP parameters of $B^0_s$--$\bar B^0_s$ mixing allows us to take these effects into account in this determination of the $B^0_s\to \mu^+\mu^-$ branching ratio, as studied in detail in Ref.\ \cite{Bq-NP-analysis}.

The strategy presented in Ref.~\cite{BdBFM} provides also interesting insights into strong interaction physics, in particular factorisation. 
Having the hadronic penguin parameters at hand and using information from semileptonic $B$ decays to minimise the impact of hadronic form factors, 
effective colour-suppression factors $a_2$ can be determined from the data, showing agreement with the picture of naive 
factorisation. This is an interesting finding as factorisation is -- a priori -- not expected to work well in these colour-suppressed decays. Using these results, non-factorisable $SU(3)$-breaking effects can be constrained at the $5\%$ level, thereby showing that the
method to control the penguin effects through control channels illustrated in Fig.~\ref{fig:interplay} is indeed robust.

\boldmath
\section{The $B^0_s\to K^+K^-$, $B^0_d\to\pi^+\pi^-$ System}\label{sec:BsKK}
\unboldmath
The decays $B^0_s\to K^+K^-$ and $B^0_d\to\pi^+\pi^-$ with their CP conjugates offer another exciting laboratory to probe CP violation. 
The corresponding final states are CP eigenstates with eigenvalue $+1$. Consequently, the CP-violating asymmetries are described 
by Eq.\ (\ref{CP-asym-TD}). These decays are governed by colour-allowed tree and QCD penguin topologies. Interestingly, the $B^0_s\to K^+K^-$ channel, which originates from $\bar b\to \bar s u\bar u$ processes, is dominated by the QCD penguins and receives sizeable contributions from the colour-allowed tree topologies. On the other hand, the $B^0_d\to\pi^+\pi^-$ decay is caused by $\bar b\to \bar d u\bar u$ quark-level decays and is dominated by the 
colour-allowed tree amplitude while the penguin topologies lead to significant corrections. The two decays are related by interchanging all down and 
strange quarks. Consequently, the $U$-spin symmetry of strong interactions can be applied to derive relations between the corresponding hadronic 
matrix elements \cite{RF-BsKK,RF-BsKK-07,FK-BsKK,FJV-16,FJV-22}, in a similar way as for the $B^0_d \to J/\psi K_{\rm S}$, 
$B^0_s \to J/\psi K_{\rm S}$ system discussed in the previous section. 

Let us have a closer look at the decay amplitudes \cite{FJV-22}:  
\begin{align}
    A(B_s^0\to K^+ K^-) &= \sqrt{\epsilon}e^{i\gamma}\mathcal{C}^\prime \left[1 + \frac{1}{\epsilon}d' e^{i\theta'} e^{-i\gamma} \right]\ , \\
    A(B_d^0\to \pi^+\pi^-) &= e^{i\gamma}\mathcal{C}(1 + d e^{i\theta} e^{-i\gamma}) \ ,
\end{align}
where $\epsilon$ was introduced in Eq.\ (\ref{eps-def}), $\gamma$ is again the corresponding UT angle, and the primes were introduced to distinguish the
$\bar{b}\to \bar{s}$ transition from its $\bar{b}\to \bar{d}$ counterpart. The overall normalisation 
\begin{equation}\label{C-ampl}
    \mathcal{C} \equiv \lambda^3 A R_b \left[T+E + P^{(ut)} + PA^{(ut)}\right] 
\end{equation}
depends on $A\equiv \lambda^2|V_{cb}|$ and the side $R_b\propto |V_{ub}/V_{cb}|$ of the UT, as well as the colour-allowed tree amplitude $T$, an exchange amplitude $E$, the difference of penguin topologies with internal up and top quarks and the corresponding penguin annihilation topologies. The hadronic
 parameter
\begin{equation} \label{d-theta}
    d e^{i\theta}\equiv \frac{1}{R_b}\left[\frac{P^{(ct)} + PA^{(ct)}}{T+E+P^{(ut)}+PA^{(ut)}}\right]\ ,
\end{equation}
where $\theta$ is a CP-conserving strong phase, measures the ratio of tree to penguin contributions. Here $P^{(ct)}$ and $P^{(ct)}$ describe the differences of penguin amplitudes with internal charm and top quarks. Analogous expressions can be obtained for $\mathcal{C}^\prime$ and
$d' e^{i\theta'}$ through straightforward replacements. It should be emphasised that there is a one-to-one correspondence of decay topologies in the 
$B_s^0\to K^+ K^-$ and $B_d^0\to \pi^+\pi^-$ decays, i.e.\ no topologies have to be neglected when relating these decays to each other.
The U-spin symmetry implies the following relation \cite{RF-BsKK}:
\begin{equation}\label{d-rel}
d e^{i\theta}=d'e^{i\theta'}.
\end{equation}

The CP asymmetries have the following functional dependences on the parameters:
\begin{equation}
\mathcal{A}_{\text{CP}}^{\text{dir}}(B_d\to \pi^+\pi^-)=\mbox{fct}(d,\theta, \gamma), \quad \mathcal{A}_{\text{CP}}^{\text{mix}}(B_d\to \pi^+\pi^-)=
\mbox{fct}(d,\theta,\gamma,\phi_d)
\end{equation}\vspace*{-0.9truecm}
\begin{equation}
\mathcal{A}_{\text{CP}}^{\text{dir}}(B_s\to K^+K^-)=\mbox{fct}(d',\theta', \gamma), \quad \mathcal{A}_{\text{CP}}^{\text{mix}}(B_s\to K^+K^-)=
\mbox{fct}(d',\theta',\gamma,\phi_s).
\end{equation}
Since the CP-violating mixing phase $\phi_d$ and $\phi_s$ can be determined separately as we discussed in Section~\ref{sec:BJpsi}, we may use the direct and mixing-induced CP asymmetries of $B_d\to \pi^+\pi^-$ to determine $d$ as function of $\theta$. In analogy, $d'$ can be determined as function of $\theta'$ from the CP-violating observables of the $B_s\to K^+K^-$ channel. Using the relation $d=d'$ described by Eq.\ (\ref{d-rel}), the intersection of the corresponding contours allows the determination of $\gamma$ and $\theta$ as well as $\theta'$, thereby offering also an internal test of the $U$-spin symmetry \cite{RF-BsKK}. This determination of $\gamma$ is particularly interesting in view of the significant penguin contributions, which are loop 
processes and hence may well be affected by contributions from physics beyond the SM.

\begin{figure}[t] 
   \centering
    \includegraphics[width=7.5truecm]{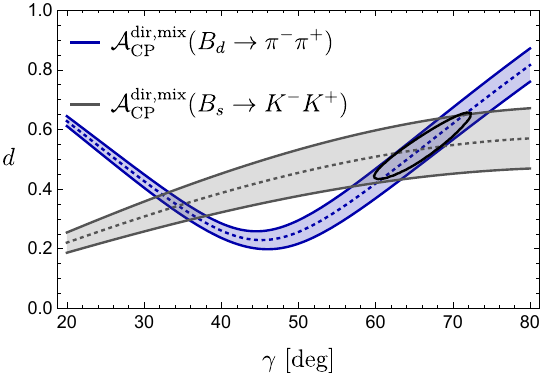} 
   \caption{Contours in the $\gamma$--$d$ plane (with $d=d'$) following from measurements of the CP-violating asymmetries of the 
   $B_d\to \pi^+\pi^-$ and $B_s\to K^+K^-$ decays (from Ref.\ \cite{FJV-22}).}\label{fig:Bs-cont}
\end{figure}

Until the recent LHCb measurement of these CP asymmetries \cite{LHCb-CP-Bs}, $\gamma$ could only be determined from this system with the help of additional information on the ratio of the branching ratios, which involves theoretical uncertainties due to hadronic effects and form factors 
\cite{RF-BsKK,RF-BsKK-07,FK-BsKK,FJV-16}. In Ref.~\cite{FJV-22}, a detailed analysis of the new LHCb results was presented. Assuming
exact $U$-spin symmetry and including also the observable $\mathcal{A}_{\rm CP}^{\Delta\Gamma}(B_s \to K^-K^+)$  (see Eq.\ (\ref{CP-asym-TD}))
in a $\chi^2$-fit to the LHCb measurements yields
\begin{equation}\label{gam-1}
    d = d' = 0.52_{-0.09}^{+0.13} \quad\mbox{with}\quad \gamma = (65_{-5}^{+7})^\circ.
    \end{equation}
The corresponding contours in the $\gamma$--$d$ plane are shown in Fig.\ \ref{fig:Bs-cont}.
Allowing for $U$-spin breaking corrections of $20\%$ through $\xi \equiv d'/d=1\pm 0.2$ yields
\begin{equation}\label{gam-U-spin}
    \gamma_{U{\text{-spin}}} = (65^{+11}_{-7})^\circ \ .
\end{equation}
We observe that the uncertainty has increased by a factor $1.5$ with respect to (\ref{gam-1}).  Using further data, the $U$-spin-breaking corrections could 
be narrowed.

Finally, also the CP-conserving strong phases
\begin{equation}
    \theta = \left(147_{-10}^{+7}\right)^\circ \ , \quad \quad \theta'= \left(114_{-10}^{+9}\right)^\circ \ 
\end{equation}
can be determined, yielding the difference
\begin{equation}
    \Delta \equiv \theta'-\theta = \left(-33 ^{+11}_{-14}\right)^\circ\ ,
\end{equation}
which would vanish in the $U$-spin limit. The $U$-spin-breaking corrections are found at the $20\%$ level. As these strong phases originate from non-factorisable processes, $U$-spin-breaking corrections at this level are not unexpected.

These are the first results using only CP violation in the $B_s\to K^+K^-$, $B_d\to \pi^+\pi^-$ system. The key question is the 
comparison with other $\gamma$ measurements. Here pure tree decays of the kind $B\to D K$ provide important strategies, as was first noticed for the 
charged $B^\pm \to D K^\pm$ modes \cite{gw,ADS,GGSZ}. Their $B^\pm_c\to D D^\pm$ counterparts would offer an ideal setting to determine 
$\gamma$ from the theoretical point of view \cite{fw}, but are very challenging experimentally. Considering only time-integrated measurements, i.e.\
excluding $B_s^0$ modes, which will be discussed in Section~\ref{sec:BsDsK}, we have the following result in a recent LHCb analysis \cite{LHCb:2021dcr}:
\begin{equation}\label{eq:gamma}
    \gamma_{B\to D K}= (64.9 \pm 4.5)^\circ \, .
\end{equation}
This determination agrees impressively with Eq.\ (\ref{gam-U-spin}). The angle $\gamma$ can also be extracted through an isospin analysis of 
$B\to \pi\pi, \rho \pi, \rho\rho$ decays employing $\phi_d$ as input. As discussed in detail in Ref.\   \cite{Bq-NP-analysis}, using 
$\phi_d$ in Eq.\ (\ref{phi_d-res}) yields
\begin{equation}\label{gamma-iso}
    \gamma_{\rm iso} = (72.6^{+4.3}_{-4.9})^\circ \ .
\end{equation}
It is remarkable that these three determinations are consistent with one another at the $1\,\sigma$ level, where the results in Eqs.\ (\ref{gam-U-spin}) and 
(\ref{gamma-iso}) could in particular be affected by NP effects entering through QCD penguin topologies. 

In Ref.\ \cite{LHCb-CP-Bs}, LHCb has also reported new measurements for the CP asymmetries of the $B_d^0 \to \pi^- K^+$ and $B_s^0 \to K^- \pi^+$ decays. These channels are related through the $U$-spin symmetry to each other as well \cite{GR-BpiK,RF-BsKK-07,FK-BsKK}. Since their final states are flavour-specific, they do not show mixing-induced CP violation. The new LHCb results show interesting patterns, as pointed out and analysed 
in Ref.\ \cite{FJV-22}: 
\begin{equation}\label{diff-1}
	\mathcal{A}_{\rm CP}^{\rm dir}(B_s^0 \to K^- K^+) - \mathcal{A}_{\rm CP}^{\rm dir}(B_d^0 \to \pi^- K^+) = 0.089 \pm 0.031 \, ,
\end{equation}
which differs from zero by $2.9 \,\sigma$. The $B_d^0 \to \pi^- \pi^+$ and $B_s^0 \to K^- \pi^+$ modes show a similar feature:
\begin{equation}\label{diff-2}
	\mathcal{A}_{\rm CP}^{\rm dir}(B_d^0 \to \pi^- \pi^+) - \mathcal{A}_{\rm CP}^{\rm dir}(B_s^0 \to K^- \pi^+) = -0.095 \pm 0.040 \, . 
\end{equation}
The corresponding decays differ only by their spectator quarks and exchange ($E$) and penguin annihilation ($PA$) topologies which enter the 
$B^0_s\to K^+K^-$, $B^0_d\to \pi^+\pi^-$ system, as we have already noted in Eqs.\ (\ref{C-ampl}) and (\ref{d-theta}), but are not present in the
$B_d^0 \to \pi^- K^+$, $B_s^0 \to K^- \pi^+$ system. These topologies are expected to play a minor role, and are usually neglected. Within this 
approximation, the direct CP asymmetries of these decays would be equal. In fact, the difference in the CP asymmetries in Eqs.\ (\ref{diff-1}) and (\ref{diff-2})
is challenging to explain through NP effects since the decays originate from the same quark-level processes. The analysis in Ref.\ \cite{FJV-22} has shown
that the differences of the CP asymmetries can actually be accommodated within the SM through exchange and penguin-annihilation contributions of reasonable size, i.e.\ not showing any anomalous enhancement. Since these topologies are highly non-factorizable, they cannot be reliably computed and have to be determined from data. A strategy was developed in that paper to constrain and determine the $E$ and $PA$ contributions from experimental 
data using again the $U$-spin symmetry. Important control channels are the $B^0_s\to \pi^+\pi^-$, $B^0_d\to K^+K^-$ decays, which offer also CP-violating asymmetries. 

The CP asymmetries of the $B^0_s\to K^+K^-$ channel can also be used to determine the $B^0_s$--$\bar B^0_s$ mixing phase $\phi_s$ 
\cite{FJV-16,FJV-22}, thereby providing an interesting alternative to the $B^0_s\to J/\psi\phi$ modes discussed in Section~\ref{sec:BJpsi}. In particular NP entering at the decay amplitude level may affect these modes differently, thereby resulting in different values of $\phi_s$. Moreover, the $\phi_s$ determination from $B^0_s\to K^+K^-$ can be performed in way that is particularly robust with respect to $U$-spin-breaking effects. The starting point is the following 
counterpart of Eq.\ (\ref{CP-rel-1}):
\begin{equation}
    \sin\phi_s^{\rm eff} = \frac{ \mathcal{A}_{\rm CP}^{\rm mix}(B_s^0\to K^-K^+)}{ \sqrt{1-\left[\mathcal{A}_{\rm CP}^{\rm dir}(B_s^0\to K^-K^+)\right]^2}} 
\end{equation}
with
\begin{equation}\label{phi_sKK}
    \phi_s^{\rm eff} \equiv \phi_s + \Delta \phi_{KK} \ ,
\end{equation}
where $\Delta\phi_{KK}$ a hadronic phase shift which can be expressed in terms of $d' $, $\theta'$ and $\gamma$. The LHCb measurements in 
Ref.\ \cite{LHCb-CP-Bs} result in 
\begin{equation}\label{phis-eff}
	\phi_s^{\rm eff} = -(8.1 \pm 1.9)^\circ,
\end{equation}
which has an impressively small uncertainty. 

In Ref.\ \cite{FJV-16}, a new strategy was developed to determine $\Delta\phi_{KK}$ in an optimal way with the help of semileptonic decays. 
One of the key aspects is the use of (double) ratios of non-leptonic and semileptonic $B_{(s)}$ decay rates: 
\begin{equation}\label{def-Rpi}
	R_{\pi}  \equiv  \frac{ \Gamma(B_d^0\to \pi^-\pi^+)}{|d\Gamma(B^0_d\rightarrow \pi^- \ell^+ 
	\nu_\ell)/dq^2|_{q^2=m_\pi^2}} = 6\pi^2 (|V_{ud}| f_\pi)^2 X_{\pi} |a_{\rm NF}|^2 r_\pi ,
\end{equation}
where $f_\pi$ is the pion decay constant  and $X_{\pi}$ a ratio of phase-space and form factors:
\begin{equation}\label{eq:Xdef}
    X_\pi \equiv \frac{(m_{B_d}^2 - m_\pi^2)^2}{m_{B_d}^2 ( m_{B_d}^2-4m_\pi^2)}  \left[ \frac{F_0^{B_d\pi}(m_\pi^2)}{F_1^{B_d\pi}(m_\pi^2)} \right]^2.
\end{equation}
For the definition of the form factors $F_0$ and $F_1$, see Ref.\ \cite{FJV-16}. Interestingly, their ratio is given exactly by one at $q^2 = 0$ due to kinematic 
constraints. Since for $q^2=m_\pi^2$ we are close to this situation, the form factor dependence essentially drops out in $X_\pi$. The $a_{\rm NF}$ describes
non-factorisable effects, and the hadronic parameters enter through
\begin{equation}\label{eq:defrpi}
r_\pi \equiv 1-2d\cos\theta\cos\gamma	+d^2	 \ .
\end{equation}
Using $\phi_d$ and $\gamma$ as input parameters from Eqs.\ (\ref{phi_d-res}) and (\ref{eq:gamma}), respectively, allows us to extract the penguin parameters $d$ and $\theta$ from the CP asymmetries of $B_d^0 \to \pi^- \pi^+$, thereby providing a theoretically clean determination of $r_\pi$.

In analogy to $R_\pi$ in (\ref{def-Rpi}),  a ratio $R_K$ can be introduced for $B^0_s\to K^+K^-$ involving the $B^0_s\rightarrow K^- \ell^+ \nu_\ell$ 
decay \cite{FJV-16}. Taking the ratio of $R_\pi$ and ${R}_K$ yields
\begin{equation} \label{rkTh}
r_K    = \frac{{R}_K}{R_\pi} \left( \frac{|V_{ud}| f_\pi}{|V_{us}| f_K} \right)^2 \frac{X_\pi}{{X}_K} \left( {\xi}_\text{NF}^a \right)^2 r_\pi ,
\end{equation}
allowing us to determine
\begin{equation}\label{rK-expr}
r_K \equiv 1+2 \frac{d^\prime}{\epsilon} \cos\theta^\prime \cos\gamma + \left(\frac{d^{\prime}}{\epsilon}\right)^2.
\end{equation}
The ratio of the CKM factors and pion and kaon decay constants in Eq.\ (\ref{rkTh}) can be determined with high precisoin from data \cite{PDG}. 
Consequently, the only remaining theoretical uncertainty enters through
\begin{equation}\label{eq:xidef}
    {\xi}_\text{NF}^a \equiv \left| \frac{1 + r_P}{1 + {r}'_P} \right| \left| \frac{1 + x}{1+x'} \right| \left| \frac{a_{\rm NF}^T}{{a}_{\rm NF}^{T'}} \right|,
\end{equation}
which parametrises the non-factorisable $U$-spin-breaking contributions. In the limit of exact $U$-spin symmetry, we have $\xi_{\rm NF}^a=1$. Thanks to the use of the semileptonic ratios, the non-factorisable effects enter only in the form of double ratios. This leaves a very favourable structure from the perspective of potential $U$-spin-breaking corrections since these effects do not enter linearly \cite{FJV-16}. A detailed discussion using the most recent data is given in 
Ref.\ \cite{FJV-22}, finding 
\begin{equation}\label{xi_NFa}
\xi_{\rm NF}^a = 1.00 \pm 0.07.
\end{equation}

Using the experimental $R_K/R_\pi$ ratio, we may finally determine $r_K$, allowing us to determine $d'$ as a function of $\theta'$ through 
Eq.~(\ref{rK-expr}). 
Another function of $d'$ and $\theta'$ is provided by the direct CP asymmetry of $B_s^0 \to K^-K^+$. Consequently, we have sufficient information to extract $d'$ and $\theta'$, thereby allowing us to calculate $\Delta\phi_{KK}$. Finally, using this hadronic phase shift,  the effective mixing phase $\phi_s^{\rm eff}$ in
Eq.\ (\ref{phi_sKK}) can be converted into the $B_s^0$--$\bar{B}_s^0$ mixing phase $\phi_s$.  

The LHCb collaboration has observed the semileptonic $B_s^0 \to K^- \ell^+ \nu_\ell$ decay \cite{LHCb:2020ist}. Although the 
integrated rate is measured in different regions of $q^2$, results for the differential rate at $q^2=m_K^2$ have unfortunately not yet been reported. Therefore $R_K$ and hence $\Delta \phi_{KK}$ cannot yet be determined with this strategy. Once available, this method would be the most favourable to pursue because any experimental improvement directly leads to a more precise determination of $\phi_s$. In Ref.\ \cite{FJV-16}, it was discussed in detail how the theoretical uncertainty, i.e. the uncertainty on $\xi_{\rm NF}^a$, compares to the experimental uncertainties. Taking the range for $\xi_{\rm NP}^a$ in Eq.\ (\ref{xi_NFa}) gives a theoretical uncertainty of only $0.8^\circ$ for $\Delta\phi_{KK}$.

In Ref.\ \cite{FJV-22}, alternative methods are discussed and applied to circumvent the missing measurement of the differential 
$B_s^0 \to K^- \ell^+ \nu_\ell$ decay rate, yielding the result
\begin{equation} \label{phis-from-K}
    \phi_s = -(3.6 \pm 5.7)^\circ \, ,
\end{equation}
which should be compared with the value of $\phi_s$ in Eq.\ (\ref{BpsiV}) following from the analysis of CP violation in $B^0_s\to J/\psi \phi$ modes. It is interesting to note that we find remarkable agreement. However, the current uncertainties still leave significant room for physics from beyond the SM 
that can be further explored in the future. 

\boldmath
\section{The $B^0_s\to D_s^\mp K^\pm$ Decays}\label{sec:BsDsK}
\unboldmath
Another important laboratory to explore CP violation is given by the $\bar{B}^0_s\to D_s^+K^-$ and $B^0_s\to D_s^+K^-$ decays 
\cite{ADK,RF-BsDsK,dBFKMST}. In contrast to the channels discussed in the previous sections, these modes do not receive penguin contributions. 
They are governed by colour-allowed tree topologies and have also exchange contributions, which play a minor role. In contrast to the 
$B_{d(s)}^0\to J/\psi K_{\rm S}$ and $B^0_d\to K^+K^-$, $B^0_d\to \pi^+\pi^-$ modes, we have now final states which are not eigenstates of the 
CP operator. However, since both $\bar{B}^0_s$ and $B^0_s$ mesons may decay into the same final state $D_s^+K^-$, the 
$B^0_s$--$\bar{B}^0_s$ oscillations may again induce mixing-induced CP violation. The corresponding time-dependent rate asymmetry 
is given as follows:
\begin{equation}\label{BsDsK-CP}
\frac{\Gamma(B^0_s(t)\to D_s^{+} K^-) - \Gamma(\bar{B}^0_s(t)\to D_s^{+} K^-) }
	{\Gamma(B^0_s(t)\to D_s^{+} K^-) + \Gamma(\bar{B}^0_s(t)\to D_s^{+} K^-) }  
	= \frac{{C}\,\cos(\Delta M_s\,t) + {S}\,\sin(\Delta M_s\,t)}
	{\cosh(y_s\,t/\tau_{B_s}) + {\cal A}_{\Delta\Gamma}\,\sinh(y_s\,t/\tau_{B_s})}.
\end{equation}
This form is analogous to Eq.\ (\ref{CP-asym-TD}) for decays into CP eigenstates. A similar expression, with 
observables $\overline{C}$, $ \overline{S}$ and $\overline{{\cal A}}_{\Delta\Gamma}$, 
holds with straightforward replacements for the $B_s$ decays into the CP-conjugate final state $D_s^-K^+$. The quantities
\begin{equation}
 C=\frac{1-|\xi|^2}{1+|\xi|^2},  \quad S= \frac{2\,{\rm Im}{\,\xi}}{1 + |\xi|^2}, \quad 
 \mathcal{A}_{\Delta \Gamma}=\frac{2\,{\rm Re}\,\xi}{1+|\xi|^2}
\end{equation}
and their CP conjugates can be extracted from the time-dependent rate asymmetries, thereby allowing the determination of the observables 
$\xi$ and $\overline{\xi}$. In their product, the hadronic parameters cancel \cite{RF-BsDsK}:
\begin{equation}\label{xi-prod-SM}
{\xi} \times \bar{\xi}= e^{-i2( \phi_s + \gamma)} .
\end{equation}
Consequently, the CP-violating phase $\phi_s+\gamma$ can be determined in a theoretically clean way \cite{ADK,RF-BsDsK}. 
Since $\phi_s$ is determined through $B^0_s\to J/\psi \phi$ and similar modes, as we have discussed in Section~\ref{sec:BJpsi}, the UT 
angle $\gamma$ can be extracted.  

The LHCb collaboration has reported an experimental analysis of CP violation in
the $B^0_s\to D_s^\mp K^\pm$ system in Ref.~\cite{LHCb-BsDsK}, finding the result $\gamma=\left(128^{+17}_{-22}\right)^\circ$ (mod $180^\circ$). 
Here the SM relation 
\begin{equation}
C+\overline{C}=0
\end{equation}
was assumed. The result for $\gamma$ is puzzling since analyses of the UT and other $\gamma$ 
determinations using pure tree decays give values in the $70^\circ$ regime \cite{CKMfitter, UTfit,LHCb:2021dcr}, 
as we have seen in Section~\ref{sec:BsKK}.

This intriguing situation has recently been studied in detail in Refs.~\cite{FM-1,FM-2}. Using
\begin{equation}
\tan(\phi_s+\gamma)=-\left[\frac{\overline{S}+ S}{ \mathcal{\overline{A}}_{\Delta \Gamma}+ \mathcal{A}_{\Delta \Gamma}}\right], 
\quad
\tan\delta_s= \left[\frac{\overline{S} - S}{ \mathcal{\overline{A}}_{\Delta \Gamma}+ \mathcal{A}_{\Delta \Gamma}}\right],
\end{equation}
where $\delta_s$ is the CP-conserving strong phase difference between the $\bar{B}^0_s\to D_s^+K^-$ and $B^0_s\to D_s^+K^-$ decay amplitudes, a transparent determination of these parameters is possible. The corresponding analysis gives a picture in full agreement with the complex LHCb fit. The solutions modulo $180^\circ$ can actually be excluded, since the corresponding strong phase $\delta_s$ around $180^\circ$ would be in conflict with factorisation, while $\delta_s= (-2^{+13}_{-14})^\circ$ is in excellent agreement with this
theoretical framework.

How could NP effects enter this measurement? They could give rise to new CP-violating contributions to $B^0_s$--$\bar{B}^0_s$ mixing, 
thereby affecting $\phi_s$, as we have seen in Section~\ref{sec:BJpsi}. However, such effects are included as this phase is determined through 
$B^0_s\to J/\psi \phi$ and penguin control modes. Using the corresponding value in Eq.\ (\ref{BpsiV}) taking penguin corrections into account shifts 
the LHCb result to $\gamma=\left(131^{+17}_{-22}\right)^\circ$. 

Consequently, this puzzling value of $\gamma$ would require NP contributions -- with new sources of CP violation -- at the 
decay amplitude level of the $B^0_s\to D_s^\mp K^\pm$ system. Such effects should manifest themselves also 
in the branching ratios of the corresponding decays. Concerning the branching ratios of $B_s$ decays, there are subtleties due to 
$B^0_s$--$\bar{B}^0_s$ mixing \cite{dBFKMST,BR-paper}. The ``theoretical" branching ratios refer to a situation where the
mixing effects are ``switched off":
\begin{equation}
\mathcal{B}_{\rm{th}} \equiv \frac{1}{2}\left[ \mathcal{B}(\bar{B}^0_s \rightarrow D_s^+ K^-)_{\rm{th}} +  
\mathcal{B}({B}^0_s \rightarrow D_s^+ K^-)_{\rm{th}}\right].
\end{equation}
The observable $\xi$ allows us to disentangle the decay paths (in analogy for $D_s^-K^+$):
\begin{equation}
\mathcal{B}(\bar B^0_s\to D_s^+K^-)_{\rm{th}}=2 \left(\frac{|\xi|^2}{1+|\xi|^2} \right)\mathcal{B}_{\rm{th}}, \quad
\mathcal{B}(B^0_s\to D_s^+K^-)_{\rm{th}}=2 \left(\frac{1}{1+|\xi|^2} \right)\mathcal{B}_{\rm{th}}.
\end{equation}
The ``experimental" branching ratios refer to the following time-integrated rates:
\begin{equation}
\mathcal{B}_{\rm {exp}} = 
 \frac{1}{2} \int_0^{\infty} \! \left[\Gamma (\bar{B}_s^0 (t)\rightarrow D_s^+ K^-) + \Gamma (B_s^0 (t)\rightarrow D_s^+ K^-) \right]
 \mathrm{d}t ,
\end{equation}
and are related to the theoretical branching ratios through
\begin{equation}
\mathcal{B}_{\rm{th}} = \left[ \frac{1-y_s^2}{1+ \mathcal{A}_{\Delta \Gamma_s} y_s} \right] \mathcal{B}_{\rm{exp}}.
\end{equation}
Unfortunately, only a measurement of the following average is available:
\begin{equation}
\mathcal{B}^{\rm{exp}}_\Sigma \equiv \mathcal{B}_{\rm{exp}} + \bar{\mathcal{B}}_{\rm{exp}} \equiv 2\, \langle\mathcal{B}_{\rm{exp}}\rangle = (2.27 \pm 0.19) \times 10^{-4}.
\end{equation}
Assuming the SM, as the LHCb collaboration, yields
\begin{equation}
 \mathcal{B}_{\rm{th}} =  \bar{\mathcal{B}}_{\rm{th}}=
\left[\frac{1-y_s^2}{1+y_s\langle {\cal A}_{\Delta\Gamma}\rangle_+}\right]\langle\mathcal{B}_{\rm{exp}}\rangle \quad\mbox{with}\quad
\langle \mathcal{A}_{\Delta \Gamma} \rangle_+\equiv
\frac{\mathcal{\overline{A}}_{\Delta \Gamma}+ \mathcal{A}_{\Delta \Gamma}}{2}.
\end{equation}
Finally, the following branching ratios can be extracted from the data:
\begin{eqnarray}
\mathcal{B}(\bar{B}^0_s \rightarrow D_s^{+}K^{-})_{\rm th}&=&(1.94 \pm 0.21) \times 10^{-4} \\
 \mathcal{B}(B^0_s \rightarrow D_s^{+}K^{-})_{\rm th}&=&(0.26 \pm 0.12) \times 10^{-4}.
\end{eqnarray}

\begin{figure}[t!]
	\centering
\includegraphics[width = 0.48\linewidth]{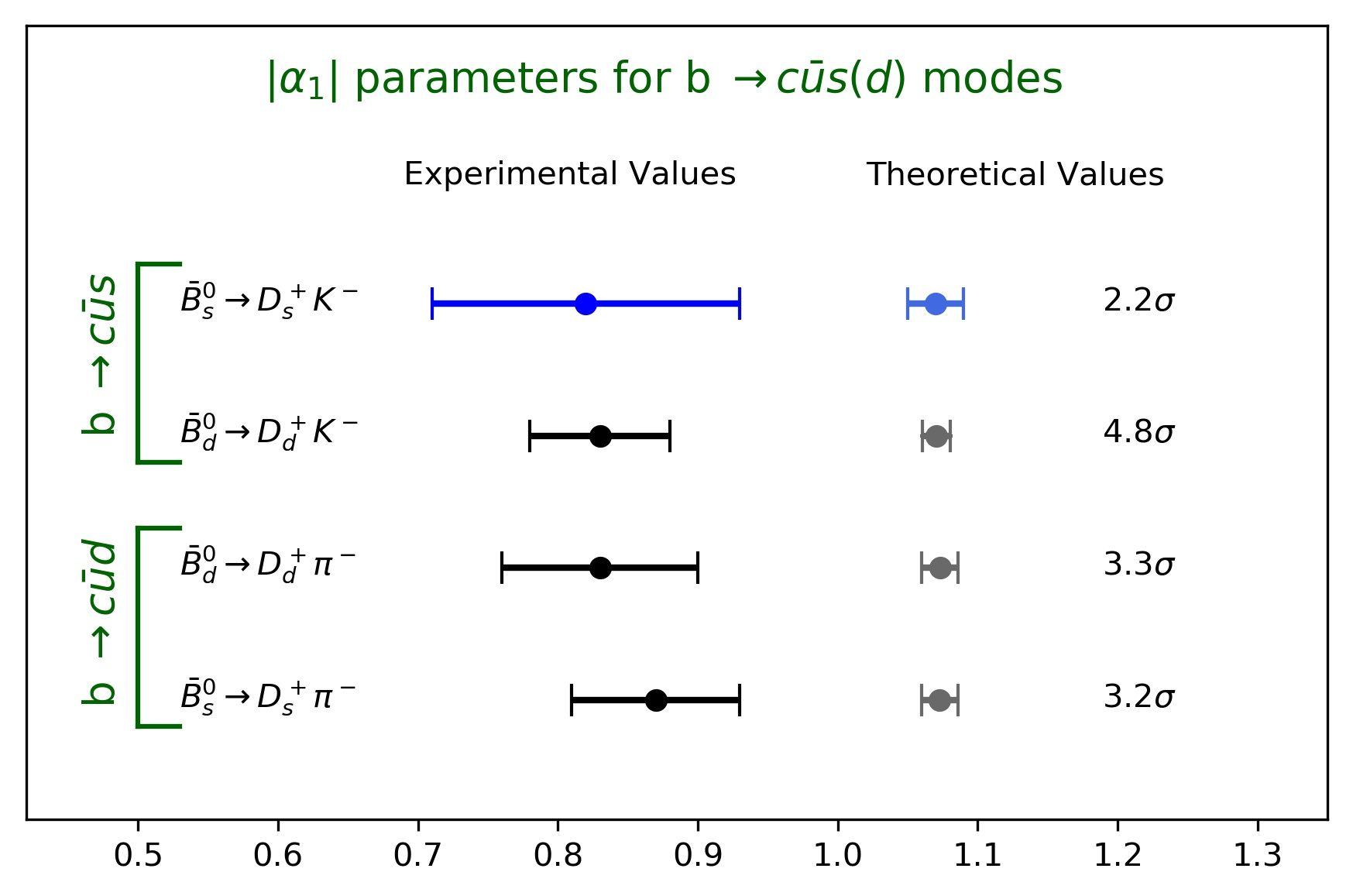}
\includegraphics[width = 0.48\linewidth]{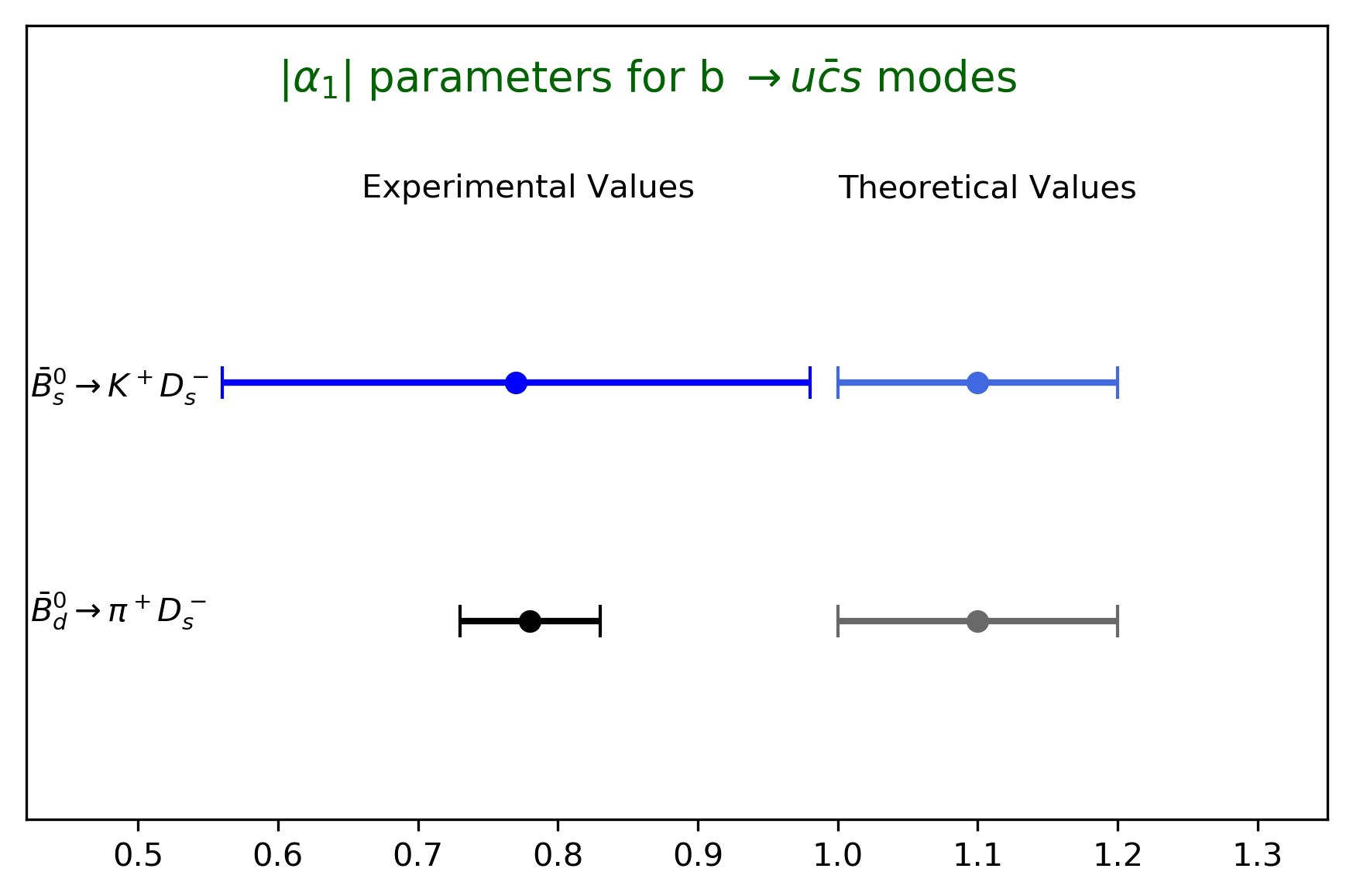}
	\caption{Comparison of the experimental and theoretical SM values of the $|a_1|$ parameters for various decay processes. 
	In the left panel, the results for $B_{(s)}$ decays caused by $b\to c \bar u s$ and $b\to c \bar u d$ processes are shown, while the right panel 
	shows the results for $b\to u\bar c s$ channels (from Ref.\ \cite{FM-2}).} \label{fig:a1}
\end{figure}

The framework for the interpretation of these quantities is provided by factorisation, which is well supported 
through the measured $\delta_s$. In order to minimise the impact of
hadronic form factor uncertainties, it is useful to introduce -- in analogy to Eq.\ (\ref{def-Rpi}) --  ratios with respect 
to semileptonic decays \cite{FM-1,FM-2}:
\begin{equation}\label{RDsK}
  R_{D_s^{+}K^{-}}\equiv\frac{\mathcal{B}(\bar{B}^0_s \rightarrow D_s^{+}K^{-})_{\rm th}}{{\mathrm{d}\mathcal{B}\left(\bar{B}^0_s \rightarrow D_s^{+}\ell^{-} \bar{\nu}_{\ell} \right)/{\mathrm{d}q^2}}|_{q^2=m_{K}^2}} = 
  6 \pi^2 f_{K}^2 |V_{us}|^2   X_{D_s K} |a_{\rm 1 \, eff }^{D_s K}|^2.
\end{equation}
Here $X_{D_s K}$ is a calculable quantity governed by phase-space effects. The parameter
\begin{equation}
a_{\rm 1 \, eff }^{D_s K}=a_{1}^{D_s K} \left(1+\frac{E_{D_s K}}{T_{D_s K}}\right)
\end{equation}
can be determined in an essentially clean way with the help of the $R_{D_s^{+}K^{-}}$ ratio. Here $a_{1}^{D_s K}$ characterises factorisation 
of the colour-allowed tree amplitude $T_{D_s K}$, which is a key application of factorisation, predicting  $|a_1^{D_sK}| = 1.07\pm0.02$ \cite{Beneke:2000ry,Huber:2016xod,Beneke:2021jhp}. The exchange amplitude $E_{D_s K}$, which gives a non-factorisable contribution, 
can be constrained through experimental data, yielding $|1+E_{D_s K}/T_{D_s K}|=1.00\pm0.08$, as discussed in detail in Ref.\ \cite{FM-1}. 
The experimental data give $|a_{\rm 1}^{D_s K}| = 0.82 \pm 0.11$, which is significantly smaller than the QCD factorisation prediction. 
Consequently, this finding shows a tension at the decay amplitude level with respect to the SM, thereby complementing the puzzling result 
for $\gamma$.

Interestingly, a similar pattern of the $|a_1|$ parameters -- with surprisingly small values -- arises also for other $B_{(s)}$ decays with similar 
dynamics. For a compilation of the corresponding values, which were also extracted in a clean way from the data utilising semileptonic decay 
information \cite{FM-1}, see the left panel of Fig.~\ref{fig:a1}. Here the $\bar B^0_d\to D_d^+ K^-$ decay stands out, showing a discrepancy of 
$4.8 \,\sigma$. Puzzlingly small branching ratios for this channel and the $\bar B^0_d\to D_d^+ \pi^-$, $\bar B^0_s\to D_s^+ \pi^-$ modes were 
also pointed out in the literature \cite{FST-BR,Bordone:2020gao}. Recently, non-factorisable effects in such decays were analysed utilising 
light-cone sum rules \cite{PR}. Studies within scenarios of physics beyond the SM were also performed \cite{Iguro:2020ndk,Cai:2021mlt,Bordone:2021cca}. 
The interesting possibility of NP effects in non-leptonic tree-level decays of $B$ mesons was discussed in Refs.\ \cite{Brod:2014bfa,Lenz:2019lvd}. 

The $|a_1|$ parameters of the $\bar{B}^0_s \rightarrow K^+ D_s^-$ and $\bar{B}^0_d \rightarrow \pi^+ D_s^-$ decays, which originate 
from $b\to u\bar c s$ processes, can also be determined through counterparts of (\ref{RDsK}) from the data; 
$\bar{B}^0_d\to \pi^+D_s^-$ differs only through the spectator quark from 
$\bar{B}^0_s \rightarrow K^+ D_s^-$. The corresponding results are shown in the right panel of Fig.\ \ref{fig:a1}. These processes are 
again governed by colour-allowed tree topologies. However, the roles of the heavy $c$ and light $u$ quarks are interchanged, 
so that the heavy-quark arguments to prove factorisation up to tiny corrections for the $b\to c$ modes do not apply, and there may be larger 
non-factorisable effects. The current uncertainties are too large to draw further conclusions. Interestingly, the 
experimental value of the strong phase $\delta_s$, which characterises the interference between the $b\to c\bar u s$ and $b\to u\bar c s$ decay paths, is found in excellent agreement with factorisation, thereby supporting this framework also for the $b\to u\bar c s$ channel. For a detailed discussion, see 
Refs.\ \cite{FM-1,FM-2}.

\begin{figure}[t] 
   \centering
   \includegraphics[width=5.5truecm]{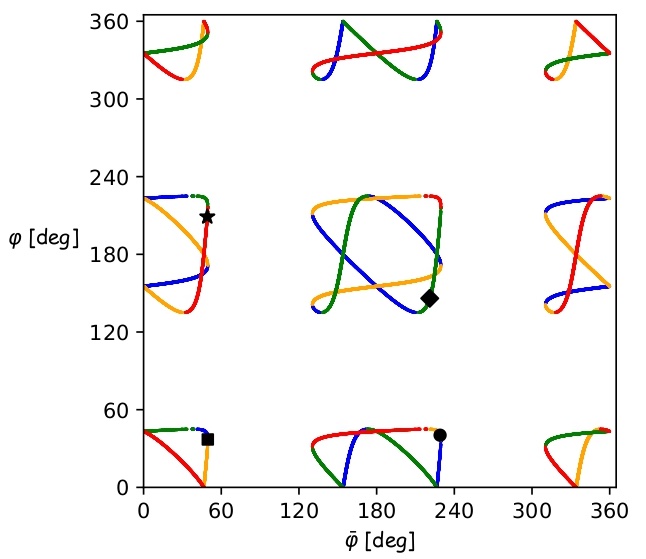}
    \includegraphics[width=5.0truecm]{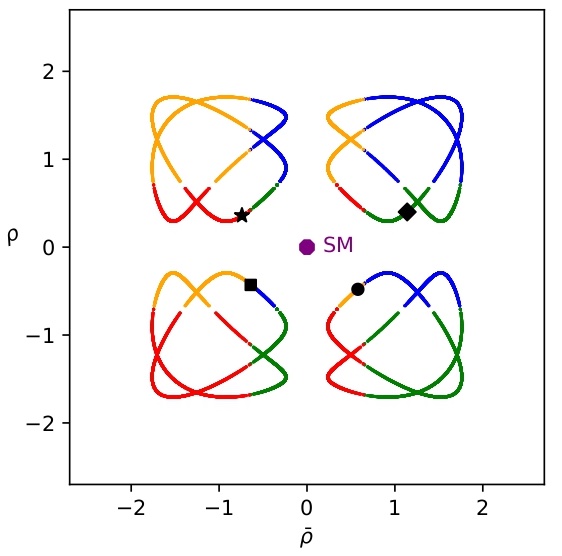} 
   \caption{Correlations between the NP parameters of the $B^0_s\to D_s^\mp K^\pm$  system (from Ref. \cite{FM-1}).}\label{fig:NP-corr}
\end{figure}

The puzzling pattern of the $|a_1|$ values for the $b\to c \bar u s$ modes complements the puzzling result for $\gamma$ from CP violation in 
$B^0_s\to D_s^\mp K^\pm$ decays in an exciting way. In order to include NP effects, the $\bar{B}^0_s \rightarrow D_s^+ K^-$ 
amplitude can be generalised as 
\begin{equation}
A(\bar{B}^0_s \rightarrow D_s^+ K^-) = A(\overline{B}^0_s \rightarrow D_s^+ K^-)_{{\rm{SM}}} \left[ 1 + \bar{\rho} \, e^{i \bar{\delta}}
e^{+i \bar{\varphi}} \right]
\end{equation}
with
\begin{equation}
\bar{\rho} \, e^{i \bar{\delta}} e^{i \bar{\varphi}}  \equiv \frac{ A(\bar{B}^0_s \rightarrow D_s^+ K^-)_{{\rm{NP}}} }{  A(\bar{B}^0_s \rightarrow D_s^+ K^-)_{{\rm{SM}}} },
\end{equation}
where $\bar{\delta}$ and $\bar{\varphi}$ are CP-conserving and CP-violating phases, respectively. An analogous expression holds for 
the $\bar{B}^0_s \rightarrow D_s^- K^+$ channel. The generalisation of the SM relation in Eq.\ (\ref{xi-prod-SM}) is given as follows:
\begin{equation}
\xi \times \bar{\xi}  = \sqrt{1-2\left[\frac{C+\bar{C}}{\left(1+C\right)\left(1+\bar{C}\right)}
\right]}e^{-i\left[2 (\phi_s +\gamma_{\rm eff})\right]}.
\end{equation}
Here $\gamma$ enters through the ``effective" angle 
\begin{equation}
\gamma_{\rm eff}\equiv \gamma+\frac{1}{2}\left(\Delta\Phi+\Delta\bar{\Phi}\right)=
\gamma-\frac{1}{2}\left(\Delta\varphi+\Delta\bar{\varphi}\right),
\end{equation}
where the NP phase shifts can be expressed in terms of $\bar{\rho}$, $\bar{\delta}$, $\bar{\varphi}$ and their CP conjugates as discussed in detail in Refs.~\cite{FM-1,FM-2}. Using the experimental information encoded in the branching ratios and CP asymmetries, the constraints on the NP parameters shown in Fig.~\ref{fig:NP-corr} can be obtained for the central values of the observables. Taking also the uncertainties into account, NP amplitudes 
in the (30-50)\% range of the SM amplitudes could accommodate the current experimental data.

Using the data collected between 2015 and 2018, the LHCb collaboration has recently presented an update for the measurement of CP violation in the 
$B^0_s\to D_s^\mp K^\pm$, reporting a value of $\gamma=(74\pm11)^\circ$ (modulo $180^\circ$) \cite{LHCb-BsDsK-update}. It will be interesting to see how the data will evolve with the full data set and in the future high-precision era.

\newpage

\boldmath
\section{Rare $B$ Decays}\label{sec:rareB}
\unboldmath
As we have seen in the previous sections, non-leptonic decays of $B$ mesons play a key role for the exploration of CP violation. However, 
in recent years, rare decays of the kind $B\to K^{(*)} \ell^+\ell^-$ have been in the spotlight in view of anomalies in certain observables \cite{ADL-rev}.
Moreover, the observation of the leptonic rare decay $B^0_s\to \mu^+\mu^-$ by the LHCb and CMS collaborations has been a highlight of 
the LHC physics programme \cite{LHCb-CMS-Bsmumu}. 
For one billion of $B^0_s$ mesons, only about three decay into a muon pair, thereby making this an incredibly 
rare process. The experimental results for the branching ratio are in the ballpark of the SM predictions \cite{Bq-NP-analysis,ADG-Bsmumu,BBS}. 
The branching ratio is so tiny due to the loop and a helicity suppression which is effective in the SM. This channel is particularly 
sensitive to probe new (pseudo)-scalar contributions. 

These leptonic and semileptonic rare $B$ decays are flavour-changing neutral current processes and hence do not arise at 
the tree level in the SM, i.e.\ require loop contributions. Consequently, they are considered as particularly sensitive probes for NP effects. The final states of
these channels are much simpler than the non-leptonic $B$ decays concerning the impact of strong interactions: In the case of $B^0_q\to \ell^+\ell^-$ decays
($q=d,s$), the hadronic binding effects are only described by the decay constant $f_{B_s}$, while in the case of semileptonic decays of the kind 
$B\to K^{(*)} \ell^+\ell^-$, the non-perturbative effects are described by form factors. These quantities can be calculated with the help of non-perturbative techniques, where lattice QCD is the key player. 

Let us first have a closer look at the leptonic $B^0_q\to \ell^+\ell^-$ decays. Before turning to CP violation in these modes \cite{CBV-Bsellell}, 
let us summarise their theoretical description. The corresponding low-energy effective Hamiltonian can be written as follows \cite{Bee}:
\begin{equation}\label{Heff}
{\cal H}_{\rm eff}=-\frac{G_{\rm F}}{\sqrt{2}\pi} V_{tq}^\ast V_{tb} \alpha
\bigl[C^{q,\ell\ell}_{10} O_{10} + C^{q,\ell\ell}_{S} O_S + C^{q,\ell\ell}_P O_P+ 
C_{10}^{q,\ell\ell'} O_{10}' + C_{S}^{q,\ell\ell'} O_S' + C_P^{q,\ell\ell'} O_P' \bigr].
\end{equation}
Here $\alpha$ is the QED coupling and the Wilson coefficients $C_i^{q,\ell\ell}$,  $C_i^{q,\ell\ell'}$ describe the short-distance 
physics of the the four-fermion operators
\begin{equation}
\begin{array}{rclcrcl}
O_{10}&=&(\bar q \gamma_\mu P_L b) (\bar\ell\gamma^\mu \gamma_5\ell), & \mbox{}\qquad &
 O_{10}'&=&(\bar q \gamma_\mu P_R b) (\bar\ell\gamma^\mu \gamma_5\ell),  \\
O_S&=&m_b (\bar q P_R b)(\bar \ell \ell), & \mbox{}\qquad &  O_S'&=&m_b (\bar q P_L b)(\bar \ell \ell),\\
O_P&=&m_b (\bar q P_R b)(\bar \ell \gamma_5 \ell), & \mbox{}\qquad  
& O_P'&=&m_b (\bar q P_L b)(\bar \ell \gamma_5 \ell),
\end{array}
\end{equation}
where $P_{L,R}\equiv(1\mp\gamma_5)/2$, $m_b$ is the $b$-quark mass, and the $O'_i$  
are obtained from the $O_i$ through the replacements $P_L \leftrightarrow P_R$.
The corresponding matrix elements can be expressed in terms of the $B_q$-meson decay constant $f_{B_q}$.
It is useful to introduce 
\begin{equation}\label{P-expr}
P^q_{\ell\ell}\equiv |P^q_{\ell\ell}|e^{i\varphi_{P_q}^{\ell\ell}} \equiv \frac{C_{10}^{q,\ell\ell}-C_{10}^{q,\ell\ell'}}{C_{10}^{\rm SM}}+\frac{M_{B_q}^2}{2 m_\ell}
\left(\frac{m_b}{m_b+m_q}\right)\left[\frac{C^{q,\ell\ell}_P-C_P^{q,\ell\ell'}}{C_{10}^{\rm SM}}\right]
\end{equation}
\begin{equation}\label{S-expr}
S^q_{\ell\ell}\equiv |S^q_{\ell\ell}|e^{i\varphi_{S_q}^{\ell\ell}} \equiv \sqrt{1-4\frac{m_\ell^2}{M_{B_q}^2}}
\frac{M_{B_q} ^2}{2 m_\ell}\left(\frac{m_b}{m_b+m_q}\right)
\left[\frac{C_S^{q,\ell\ell}-C_S^{q,\ell\ell'}}{C_{10}^{\rm SM}}\right],
\end{equation}
where $\varphi_{P_q}^{\ell\ell}$ and $\varphi_{S_q}^{\ell\ell}$ are CP-violating phases. The combinations of Wilson coefficients 
in Eqs.~(\ref{P-expr}) and (\ref{S-expr}) were introduced to have the simple SM relations 
\begin{equation}
P^q_{\ell\ell}|_{\rm SM}=1, \quad S^q_{\ell\ell}|_{\rm SM}=0.
\end{equation}
Further details can be found in Ref.\ \cite{Bee}, pointing out also possible huge enhancements of the $B^0_s\to e^+e^-$ branching ratio with 
respect to the extremely small SM value.

For our discussion of CP violation, we focus on $B^0_s\to \mu^+\mu^-$. New sources of CP violation may enter through complex phases 
of the short-distance coefficients. In analogy to Eq.~(\ref{BsDsK-CP}), we introduce the following time-dependent CP asymmetry \cite{ADG-Bsmumu,BFGK}:
\begin{equation}\label{CP-asym-Bsmumu}
\frac{\Gamma(B^0_s(t)\to \mu_\lambda^+\mu^-_\lambda)-
\Gamma(\bar B^0_s(t)\to \mu_\lambda^+\mu^-_\lambda)}{\Gamma(B^0_s(t)\to \mu_\lambda^+\mu^-_\lambda)+
\Gamma(\bar B^0_s(t)\to \mu_\lambda^+\mu^-_\lambda)}
=\frac{{\cal C}_{\mu\mu}^\lambda\cos(\Delta M_st)+{\cal S}_{\mu\mu}\sin(\Delta M_st)}{\cosh(y_st/\tau_{B_s}) + 
{\cal A}_{\Delta\Gamma_s}^{\mu\mu} \sinh(y_st/\tau_{B_s})},
\end{equation}
where $\lambda$ is the muon helicity. Introducing $P\equiv P^s_{\mu\mu}$ and  $S\equiv S^s_{\mu\mu}$ for the combinations in Eqs.~(\ref{P-expr}) 
and (\ref{S-expr}), respectively, the observables take the following forms:
\begin{equation}
{\cal C}_{\mu\mu}^\lambda 
=  -\eta_\lambda\left[\frac{2|PS|\cos(\varphi_P-\varphi_S)}{|P|^2+|S|^2} 
	\right] \equiv -\eta_\lambda {\cal C}_{\rm \mu\mu} 
\end{equation}
\begin{equation}
{\cal S}_{\mu\mu}^\lambda
	=\frac{|P|^2\sin(2\varphi_P-\phi_s^{\rm NP})-|S|^2\sin(2\varphi_S-\phi_s^{\rm NP})}{|P|^2+|S|^2}
	\equiv {\cal S}_{\mu\mu}
\end{equation}
\begin{equation}\label{ADGs}
{\cal A}^{\mu\mu}_{\Delta\Gamma_s} = \frac{|P^s_{\mu\mu}|^2\cos(2\varphi_{P_s}^{\mu\mu}-\phi_s^{\rm NP}) - 
|S^s_{\mu\mu}|^2\cos(2\varphi_{S_s}^{\mu\mu}-\phi_s^{\rm NP})}{|P^s_{\mu\mu}|^2 + |S^s_{\mu\mu}|^2},
\end{equation}
where $\phi_s^{\rm NP}$ is the NP component of the $B^0_s$--$\bar B^0_s$ mixing phase in Eq.\ (\ref{phi-q}), 
while $\eta_{\rm L}=+1$ and $\eta_{\rm R}=-1$. Unfortunately, the measurement of the muon helicity is very challenging. However, 
we may also consider the helicity-averaged rates. Here the ${\cal C}_{\mu\mu}^\lambda$ term cancels:
\begin{equation}
\frac{\Gamma(B^0_s(t)\to \mu^+\mu^-)-
\Gamma(\bar B^0_s(t)\to \mu^+\mu^-)}{\Gamma(B^0_s(t)\to \mu^+\mu^-)+
\Gamma(\bar B^0_s(t)\to \mu^+\mu^-)}
=\frac{{\cal S}_{\mu\mu}\sin(\Delta M_st)}{\cosh(y_st/ \tau_{B_s}) + 
{\cal A}_{\Delta\Gamma_s}^{\mu\mu} \sinh(y_st/ \tau_{B_s})}.
\end{equation}
In analogy to Eq.~(\ref{cons-rel}), the CP asymmetries satisfy the relation
\begin{equation}
({\cal C}_{\mu\mu}^\lambda)^2+({\cal S}_{\mu\mu})^2+({\cal A}_{\Delta\Gamma_s}^{\mu\mu})^2=1. 
\end{equation}
The allowed ranges for ${\cal S}_{\mu\mu}$ and ${\cal A}_{\Delta\Gamma_s}^{\mu\mu}$ were studied in specific NP scenarios in 
Ref.~\cite{BFGK}, while a detailed analysis to probe possible CP-violating phases of $P$ and $S$ was performed in Ref.~\cite{CBV-Bsellell}. 
In particular, the measurement of the $B_s\to \mu^+\mu^-$ observables would allow the determination of the short-distance coefficients as 
functions of the CP-violating phase $\varphi_S$. The corresponding measurements offer exciting new perspectives for the LHCb upgrade(s).
 Detailed feasibility studies would be very desirable. 

In semileptonic decays of the kind $B\to K \ell^+\ell^-$ and $B\to K^*\ell^+\ell^-$, $B_s\to \phi\ell^+\ell^-$, CP violation can also be explored 
through a variety of observables \cite{BHP,BHvDy,HS,D-GV,BFKS,D-GN-BV,FP-paper,RK-CP}. In these channels, we get contributions from 
$\bar c c$ resonances through intermediate $J/\psi$, $\psi(2S)$, ...\ states that originate from matrix elements of current--current operators. Since these are 
non-perturbative long-distance effects, the momentum transfers $q^2$ to the $\ell^+\ell^-$ pair in the resonance region are 
usually excluded in the analyses of the semileptonic rare $B$ decays. However, these contributions can be described through fits to further experimental 
data \cite{FP-paper,LHCb-model}. For the exploration of CP violation, these effects are actually very interesting as they provide CP-conserving 
phases which are -- in addition to CP-violating phases -- an essential requirement for direct CP violation \cite{BFKS,FP-paper}. In order to search for 
such CP asymmetries, which would signal new sources of CP violation, measurements of the semileptonic rare $B$ decays in the resonance region 
should be pursued and would be very interesting. 

For decays such as $B_d \to K_{\rm S}\ell^+\ell^-$, $B_d\to K^*(\to \pi^0K_{\rm S})\ell^+\ell^-$ and $B_s \to \phi \ell^+\ell^-$, we have also mixing-induced
CP-violating phenomena since both $B^0_{d(s)}$ and $\bar B^0_{d(s)}$ mesons may decay into the corresponding final states, in analogy to the 
time-dependent CP asymmetry described by Eq.\ (\ref{CP-asym-Bsmumu}). Detailed discussions of the corresponding observables can be found in 
Refs.\  \cite{BHP,D-GV,D-GN-BV}. A new strategy to extract complex Wilson coefficients with the help of direct CP violation in 
$B^+\to K^+ \mu^+\mu^-$,  $B^0_d \to K_{S} \mu^+\mu^-$ and the mixing-induced CP violation in $B^0_d \to K_{S} \mu^+\mu^-$, with information
of the differential decay rate in appropriate $q^2$ bins, was proposed and illustrated for a variety of scenarios in Ref.\ \cite{FP-paper}. Here the 
complementarity of the direct and mixing-induced CP asymmetries with respect to the impact of hadronic effects and strong phases plays a key role. 
The corresponding ``fingerprinting" can also be applied in a broader context to other decays, and offers interesting new studies for the high-precision era.

The $B\to K^{(*)} \ell^+\ell^-$ modes have received a lot of attention in recent years in view of data indicating a violation of the lepton flavour 
universality present in the SM. In particular, the data suggested that electrons and muons may couple differently to NP, resulting in different 
decay rates for the $B\to K^{(*)} \mu^+\mu^-$ and $B\to K^{(*)} e^+e^-$ channels~\cite{ADL-rev}. These effects are quantified 
by ratios of the following kind:
\begin{equation}\label{RK-def}
   \langle R_K \rangle  \equiv \frac{ \Gamma(B^-\to K^-\mu^+\mu^-) +\Gamma(B^+\to K^+\mu^+\mu^-) }{\Gamma(B^-\to K^-e^+e^-)  + \Gamma(B^+\to K^+e^+e^-) } .
\end{equation}
Here the decay rates actually refer to the $q^2 \in [1.1,6.0] \, \mbox{GeV}^2$ range below the $c\bar c$ resonance region. The $\langle R_{K^{(*)}}\rangle$
ratios are given with excellent theoretical precision by one in the SM \cite{BIP,ILNZ}. Until December 2022, the measurements of
$\langle R_{K^{(*)}}\rangle$ gave values around 0.8, deviating from the SM at the $3\,\sigma$ level and indicating a violation of the electron--muon universality.
LHCb has then reported the following new result \cite{LHCb-new-PRL,LHCb-new-PRD}:
\begin{equation}\label{RK-LHCb-new}
    \langle R_K \rangle = 0.949\pm 0.05 \ ,
\end{equation}
and a similar finding for the $\langle R_{K^*}\rangle$ ratio. The new results are now in agreement with lepton flavour universality at the  $1\,\sigma$ level. 

On the other hand, using the most recent lattice QCD results for the relevant $B\to K$ from factors \cite{BK-Lattice} and paying special attention to the
input CKM parameters \cite{Bq-NP-analysis}, the SM value of the CP-averaged $B^\pm\to K^\pm \mu^+\mu^-$ branching ratio is calculated as 
follows \cite{RK-CP}:
\begin{equation}\label{BKmumu-SM}
    \mathcal{B}(B^\pm \to K^\pm \mu^+\mu^-)|_{\rm SM}[1.1,6.0] = (1.83 \pm 0.14) \times 10^{-7} \ ,
\end{equation}
whereas the LHCb collaboration reports the following measurement \cite{LHCb-BR-SL-rare}:
\begin{equation}
    \mathcal{B}(B^\pm \to K^\pm \mu^+\mu^-) = (1.19 \pm 0.07)  \times 10^{-7} \ .
\end{equation}
We observe that this measurement is below the SM prediction, thereby indicating NP effects with a significance of $3.5\,\sigma$. 
In view of the new result for the $ \langle R_K \rangle$ ratio in (\ref{RK-LHCb-new}), at first sight, we may conclude that the underlying new interactions should not violate electron--muon universality. A detailed study of the space left by the current data was performed in Ref.\ \cite{RK-CP}.
Interestingly, allowing for new CP-violating effects, it was demonstrated that significant differences between the Wilson coefficients of the electronic and
muonic final states are still allowed. These effects are encoded in the CP-violating asymmetries of the neutral and charged $B\to K \ell^+ \ell^-$ decays, 
and may be revealed through potentially large differences between the CP asymmetries of transitions with electrons and muons in the final states. 

\section{Concluding Remarks}\label{sec:concl}
Decays of $B$ mesons offer a particularly exciting laboratory to explore CP violation and utilise this phenomenon as a probe for physics from beyond 
the SM. During the recent decades, we have seen impressive progress on the theoretical and experimental frontiers. For the future, the key goal is to further increase the precision at the LHCb upgrade(s) and Belle II. As we have discussed and illustrated with various recent studies, it will be essential to have 
critical assessments of the uncertainties and match the experimental with the theoretical precisions. Historically, the key actors of CP violation have 
been non-leptonic $B$ decays. In the future, it will be important to explore such effects also for rare leptonic and semileptonic $B$ decays. It will be 
exciting to see whether these studies will lead to surprises in the high-precision era of $B$ physics, and whether they will allow us to 
eventually establish deviations from the SM, involving also new sources of CP violation.

\section*{Acknowledgements} 
I am very grateful to my students and collaborators for all the enthusiasm and work on our projects, and would like to thank Eleftheria Malami for comments on the manuscript. Research discussed in this writeup 
has been supported by the Netherlands Organisation for Scientific Research (NWO).

\newpage

\end{document}